\documentclass[aps,prl,showpacs,amsmath,amssymb,amsfonts,superscriptaddress,longbibliography,lengthcheck,twocolumn]{revtex4-1}

\usepackage{graphicx}
\usepackage{subfigure}
\usepackage{amsthm}
\usepackage{verbatim}
\usepackage{dcolumn}
\usepackage{bm}
\usepackage{epsf}
\usepackage{color}
\usepackage[colorlinks=true,citecolor=blue,linkcolor=blue,urlcolor=blue]{hyperref}%
\usepackage{bbm,MnSymbol,mathrsfs,xfrac}   
\pdfoutput=1

\newcommand{\bra}[1]{\left\langle #1\right|}
\newcommand{\ket}[1]{\left|#1\right\rangle}

\newcommand{\ptr}[2]{\mathrm{tr_{#1}}\left\{#2\right\}}

\newcommand{\co}[1]{\cos{\left(#1\right)}}
\newcommand{\si}[1]{\sin{\left(#1\right)}}

\newcommand{\bla}{bla\\bla\\bla\\bla\\bla}

\newcommand{\mc}[1]{\mathcal{#1}}

\newcommand{\mrm}[1]{\mathrm{#1}}

\newcommand{\pro}[2]{|#1\rangle\langle#2|}

\newcommand{\dt}{{\Delta t}}

\begin{document}

 \title{Quantum Zeno effect in correlated qubits}

\author{Dominik \v{S}afr\'{a}nek}
\email{dsafrane@ucsc.edu}
\affiliation{SCIPP and Department of Physics, University of California, Santa Cruz, California 95064, USA}
\author{Sebastian Deffner}
\email{deffner@umbc.edu}
\affiliation{Department of Physics, University of Maryland Baltimore County, Baltimore, MD 21250, USA}

\date{\today}

\begin{abstract}
Near term quantum hardware promises to achieve quantum supremacy. From a quantum dynamical point of view, however, it is not unambiguously clear whether fundamental peculiarities of quantum physics permit any arbitrary speed-ups in real time. We show that an only recently unveiled property of the quantum Fisher information has profound implications for the rate of possible quantum information processing. To this end, we analyze an exemplary and pedagogical example for a quantum computer consisting of a computational qubit and a quantum memory. We find that frequent interaction between memory and device exhibit the quantum Zeno effect. In a second part, we show that the Zeno effect can be prevented by carefully designing the correlations and interaction between single elements of the quantum memory.
\end{abstract}

\maketitle


We are on the verge of a technological revolution. Over the last few years the first computational devices have become available that promise to exploit so-called ``quantum supremacy''  \cite{Sanders2017}. While big corporations such as Google and IBM, and smaller start-ups such as Rigetti and DWave, announce new developments in frequent succession, many fundamental questions still remain to be answered.

Quantum supremacy means that quantum computers will be able to achieve specific tasks exponentially faster than any classical computer \cite{Preskill2012}. In this context, \emph{faster} typically stands for \emph{fewer single qubit operations} that are necessary to obtain the computational result \cite{Papageorgiou2013}. Therefore, the computational ``speed-up''  in quantum physics is not necessarily related to physical time.

Ever since Heisenberg's original inception of an uncertainty relation between energy and time \cite{Heisenberg1927}, it has been apparent that classical notions such as ``speed'' and ``speed-up'' need to be carefully reconsidered. Following the seminal works of Mandelstam and Tamm \cite{Mandelstam1945} and Margolous and Levitin \cite{Margolus1998} it has been well-established that the rate of any quantum evolution is limited by the \emph{quantum speed limit}. In the simplest case of Hamiltonian dynamics, the minimal time a quantum system needs to evolve between orthogonal states is determined by $\tau_\mrm{QSL}= \mbox{max}\{\pi\hbar/(2\Delta E),\pi \hbar/(2E)\}$, where $\Delta E$ is the variance of the energy of the initial state and $E$ its mean energy above to the ground state. In more general situations the energy is replaced by the geometric properties of the considered quantum dynamics. For details  and a concise history of the quantum speed limit we refer to a recent review and references therein \cite{deffner2017quantum}.

The natural question arises to what extend the quantum speed limit constrains the performance of prospective quantum computers. Already before Feynman's inception of quantum computing \cite{Feynman1982}, Bremermann \cite{Bremermann1967} and Bekenstein \cite{Bekenstein1974,Bekenstein1990} realized that the quantum speed limit sets an upper bound on the rate with which information can be processed in and transmitted between quantum systems. However, it has also been realized that in purposefully designed quantum systems, correlations can be utilized to speed-up quantum dynamics \cite{Deffner2013PRL,Cimmarusti2015}.

The purpose of the present work is to carefully analyze how correlations affect the rate of quantum computation. To this end, we study a fully solvable model which is inspired by recent work in thermodynamics of information \cite{Deffner2013PRX} and quantum Maxwell demons \cite{Deffner2013}. Loosely speaking such systems consist of a device, i.e., the part of the systems that \emph{does} the computation, and a quantum hard disk consisting of a set of qubits. For such a simple and pedagogical model of a quantum computer, we compute the quantum speed limit from which we draw conclusions about the limitations on possible computations. As a first main result, we will see that in the absence of correlations in the quantum hard disk, the rate of computation is severely limited by fundamental properties of quantum dynamics. In particular, we will see that frequent interaction of computational device and hard disk leads to the \emph{quantum Zeno effect}. This means that if the computational device and the quantum hard disk ``talk to each other'' too frequently, the quantum state of the computer freezes out and that all computation is inhibited. As a second main result, we will then see that correlations between the qubits in the quantum hard disk assist in overcoming the Zeno effect, but also that there is a trade-off between the strength of the interaction, and the maximal rate with which a computation can be successfully performed.

\paragraph*{Preliminaries.}

We begin by outlining our model for a simple quantum computer, and by establishing notions and notations. The computational device is a single qubit, $\mc{Q}$, that is initially prepared in the mixed state
\begin{equation}
\label{eq:rho_ini}
\rho_0=(1-\nu)\pro{0}{0}+\nu\pro{1}{1},
\end{equation}
where $0\leq\nu\leq 1$, but $\nu$ is close to 1.

This qubit $\mc{Q}$ interacts with a stream of $N$ qubits, which in the limit of $N\rightarrow \infty$ constitute a simple example of a quantum information reservoir \cite{Deffner2013PRX,Deffner2013}. For the present purposes and for finite $N$ we call these qubits the ``quantum hard disk'' or ``quantum memory'', $\mc{M}$, which initially is assumed to be blank. More formally, any $k$th qubit of $\mc{M}$ is initially prepared in $\rho^\mc{M}_{k}=\pro{0}{0}_k$.

For the sake of simplicity and in complete analogy to minimal models of quantum Maxwell demons, we further assume that in every instant $\mc{Q}$ interacts only with a single qubit of $\mc{M}$ described by the unitary evolution $U(t)=\exp(-i H^\mc{Q}_{k} t)$. After a time interval $\dt$ the interaction between $\mc{Q}$ and the $k$th qubit of $\mc{M}$ is severed, and $\mc{Q}$ is put in interaction with the $(k+1)$st qubit.

\begin{figure}
\includegraphics[width=.48\textwidth]{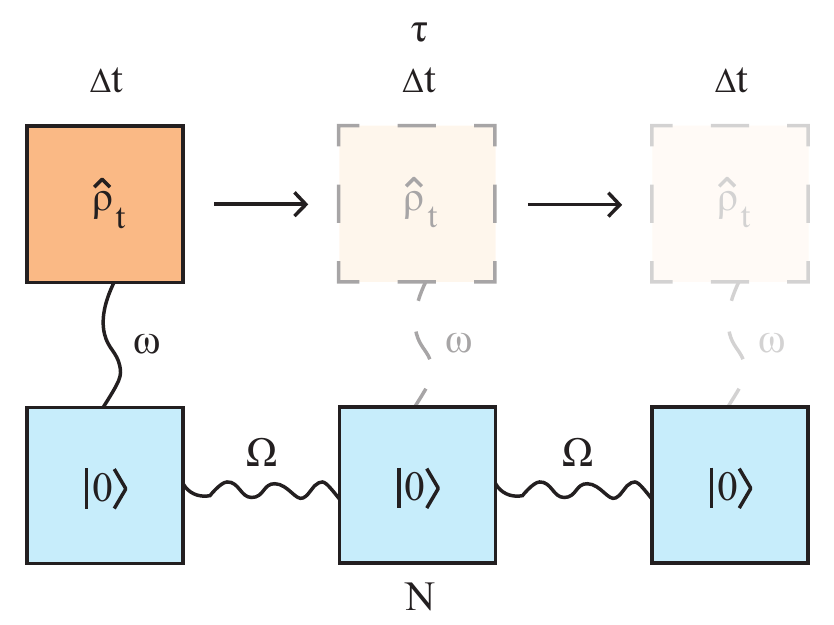}
\caption{\label{p1} Sketch of simple quantum computer: A computational qubit $\mc{Q}$ interacts with the $k$th qubit of a quantum memory $\mc{M}$ consisting of $N$ qubits for an interval of length $\Delta t$. The strength of interaction between $\mc{Q}$ and $\mc{M}$ is denoted by $\omega$, and the strength of interaction between the $N$ qubits in $\mc{M}$ is given by $\Omega$.}
\end{figure}

The interaction between $\mc{Q}$ and $\mc{M}$ is described by the Hamiltonian
\begin{equation}
\label{eq:U_t}
H^\mc{Q}_{k}=i\omega\left(\ket{0}\bra{1}_\mc{Q}\otimes\ket{1}\bra{0}_k-\ket{1}\bra{0}_\mc{Q}\otimes\ket{0}\bra{1}_k\right)\,
\end{equation}
which is a simple SWAP operation. A sketch of the system is depicted in Fig.~\ref{p1}.

Such an interaction \eqref{eq:U_t} is only capable of the simplest computation, namely partial qubit flip. After the first interval of length $\dt$, we have
\begin{equation}
\label{eq:reduced_state}
\begin{split}
\rho_t&=\ptr{1}{U(\dt)\,\rho_0\otimes\rho^\mc{M}_{1}\,U^\dag(\dt)}\\
&=\left[1-\nu\left(\cos{\dt}\right)^2\right]\pro{0}{0}+\nu\left(\cos{\dt}\right)^2\,\pro{1}{1}\,.
\end{split}
\end{equation}
Thus, we have after $N$ intervals, i.e., for times $t$ such that $N\dt\leq t\leq (N+1)\dt$,
\begin{equation}
\label{eq:reduced_state_iterations}
\begin{split}
\rho_t&=\left[1-\nu_N\,\left(\co{t-N\dt}\right)^2\right]\,\pro{0}{0}\\
&\quad+ \nu_N\,\left(\co{t-N\dt}\right)^2\,\pro{1}{1}.
\end{split}
\end{equation}
where we have introduced the \emph{purity parameter} $\nu_N=\nu\,(\cos{\dt})^{2N}$. From Eq.~\eqref{eq:reduced_state_iterations} it then becomes clear that the time to complete a full qubit flip is governed by the length of the interaction interval $\Delta t$.

\paragraph*{Quantum speed limit for qubit flips.}

Quantum dynamics can generally be characterized by the maximal rate of evolution, the quantum speed limit. This maximal rate is given by \cite{Taddei2013,deffner2017quantum},
\begin{equation}
\label{eq:vQSL}
v_\mrm{QSL}(t)=\frac{1}{2}\sqrt{F_Q(t)}\,,
\end{equation}
where $F_Q(t)$ is the quantum Fisher information.

Interestingly, $F_Q(t)$ is one of the best-studied quantities in quantum physics with a wide variety of applications. For instance, the quantum Fisher information sets bounds on the optimal estimation of parameters enconded in a quantum state~\cite{BraunsteinCaves1994a,Paris2009a,szczykulska2016multi}, it helps to describe criticality and quantum phase transitions~\cite{paraoanu1998bures,zanardi2007bures,venuti2007quantum,gu2010fidelity,banchi2014quantum,wu2016geometric,marzolino2017fisher}, it quantifies coherence and entanglement~\cite{hauke2016measuring,girolami2017information,liu2017quantum}, it provides bounds on irreversibility in open quantum systems~\cite{mancino2018geometrical}, and it also determines the best precision in thermometry \cite{Campbell2018}.

\begin{figure}
\begin{center}
\includegraphics[width=1\hsize]{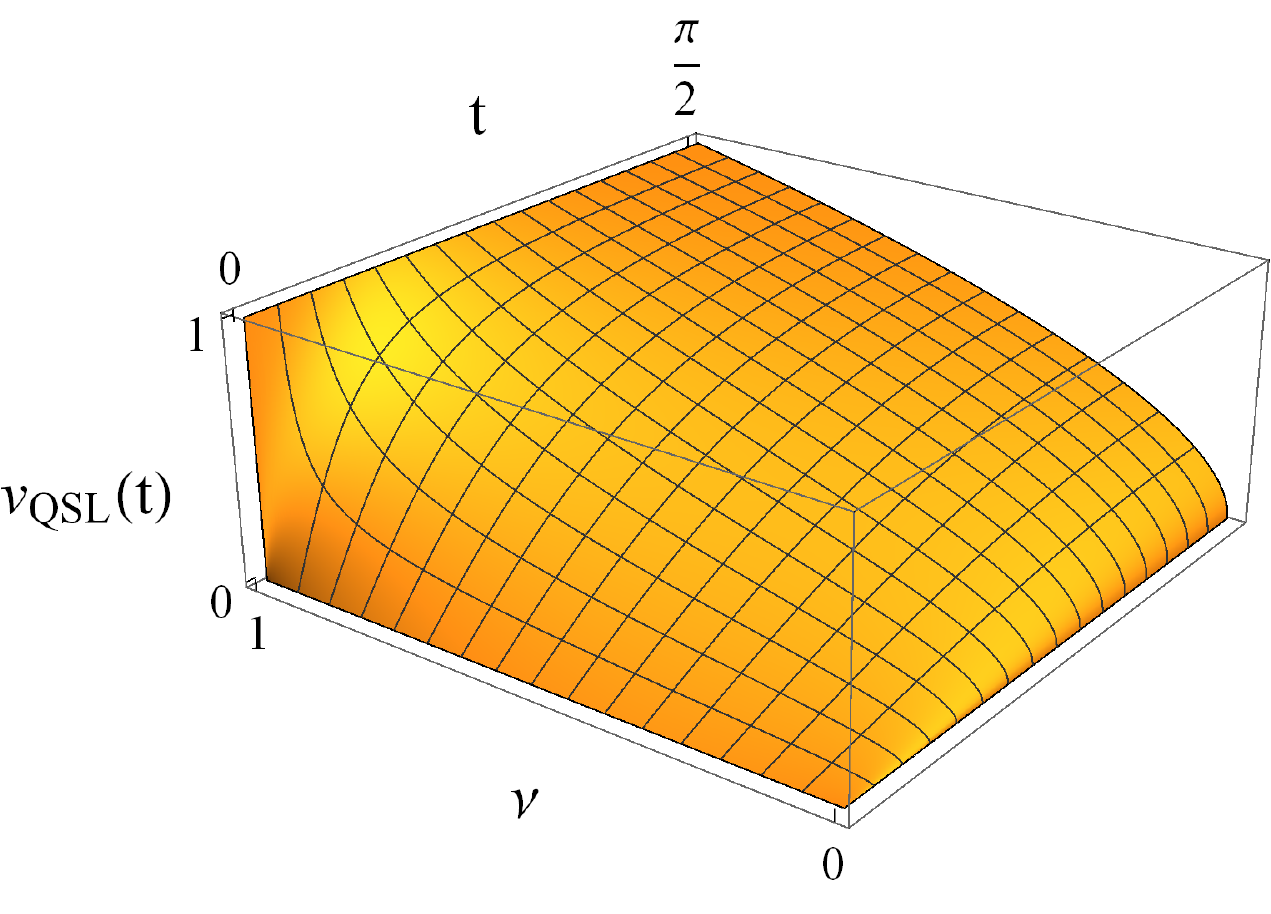}
\caption{Quantum speed limit \eqref{eq:vqsl} for the first interaction interval, that is for $t\leq \dt$. Observe the discontinuity at $(t,\nu)=(0,1)$.}
\label{fig:disc_QFI}
\end{center}
\end{figure}

Using well-known formulas from the literature \cite{Paris2009a}, the quantum speed limit \eqref{eq:vQSL} can be computed explicitly for the time-dependent state of $\mc{Q}$ \eqref{eq:reduced_state_iterations},
\begin{equation}
\label{eq:vqsl}
v_\mrm{QSL}(t)=\frac{\sqrt{\nu}\, \left|\left(\cos{\dt}\right)^N\,\si{t-N\dt} \right| }{\sqrt{1-\nu \left(\cos{\dt}\right)^{2N}\, \left[\co{t-N\dt}\right]^2}}\,.
\end{equation}
The latter expression is plotted in Fig.~\ref{fig:disc_QFI} as a function of $\nu$ and $t$. Interestingly, $v_\mrm{QSL}(t)$ \eqref{eq:vQSL} is discontinuous as a function of two variables at point $(t,\nu)=(0,1)$.  This discontinuity of the quantum Fisher information has been discussed in the literature \cite{vsafranek2017discontinuities}, however to the best of our knowledge the present analysis is the first account of relating this property with the rate of information processing.

More formally, it can be shown that the quantum Fisher information is always discontinuous at points where the rank of the density matrix changes, i.e., when subject to a purity-changing channel dependent on the parameter $t$. Since for $(t,\nu)=(0,1)$ the initial density matrix is pure, but for either $t>0$ or $\nu<1$ it is not, our situation is exactly the case where this property applies.

\paragraph*{Quantum Zeno effect from uncorrelated memories.}

The mathematical observation that $v_\mrm{QSL}(t)$ \eqref{eq:vQSL} exhibits discontinuous behavior has profound, physical implications: for $\nu$ close to $1$ and comparably small times $t$, the quantum speed limit $v_\mrm{QSL}(t)$ is very close to zero. Therefore, frequent switching between different qubits in $\mc{M}$ maintains a low rate of evolution of the reduced state of $\mc{Q}$. This switching corresponds to substituting the purity parameter $\nu_{k-1}$ for $\nu_k$ at each iteration
, and as long as $t_k=t-k\dt$ is very small comparable to $\nu_k$, $v_\mrm{QSL}(t)$ remains close to zero.

More conceptually, the discontinuous behavior of $v_\mrm{QSL}(t)$ \eqref{eq:vQSL} is a signature of the quantum Zeno effect \cite{Misra1977}. The interaction of $\mc{Q}$ with $\mc{M}$ can be understood as $\mc{M}$ taking $N$ measurements on $\mc{Q}$. If these measurement happen too frequently the quantum state of $\mc{Q}$ is prevented from evolution, and thus $\mc{Q}$ ``freezes out''. This can be made even more obvious by introducing the \emph{quantum speed limit time}, which is $v_\mrm{QSL}(t)$ averaged over an external time $\tau$,
\begin{equation}
\label{eq:tauQSLoriginal}
\tau_\mrm{QSL}= \frac{\tau}{\int_0^\tau dt\  v_\mrm{QSL}(t)}.
\end{equation}
In general, $\tau_\mrm{QSL}$ depends on $\tau$, which means that $\tau_\mrm{QSL}$ does not have a clear operational interpretation. Note, however that for pure states evolving under time-independent Hamiltonians, $\tau_\mrm{QSL}$ becomes identical to the minimal time a quantum system needs to evolve between distinguishable states \cite{deffner2017quantum}.

In the present case, where $\tau=N\dt$, we  have explicitly
\begin{equation}
\label{eq:tauQSLintegrated}
\tau_\mrm{QSL}=\frac{\tau}{\arcsin(\sqrt{\nu})-\arcsin(\sqrt{\nu}\,\left[\co{\tau/N}\right]^N)}.
\end{equation}
We observe that no matter how we choose the total time of interaction $\tau$, as long as it is fixed, $\tau_\mrm{QSL}$ diverges as $N$ goes to infinity. This means that the characteristic time it takes for the initial quantum state to evolve to an orthogonal state goes to infinity. In other words, frequent interaction stops the density matrix from evolving. We could  have immediately reached this conclusion from the reduced density matrix \eqref{eq:reduced_state}. The advantage of Eq.~\eqref{eq:tauQSLintegrated} is, however, that it gives a precise quantification of this effect.

From a practical point of view this means that in the limit of frequent interaction between $\mc{Q}$ and the quantum hard disk $\mc{M}$ successful computation is infeasible. In the remainder of this analysis we are going to show that if the qubits in $\mc{M}$ are allowed to interact and build correlations this failure of computation can be prevented.

\paragraph*{Correlated quantum hard disks.}

It has been shown that a positive feedback loop due to interaction with environments can lead to the so-called anti-Zeno effect \cite{schieve1989numerical,kofman2000acceleration,facchi2001quantum,fischer2001observation}, which means that quantum evolution can be sped up. Similarly, it has been experimentally demonstrated that quantum dynamics can experience an environment assisted speed-up in the prescence of correlations \cite{Cimmarusti2015}. Therefore, we will now show that a similar effect can be achieved for our present model by a judicious design of $\mc{M}$.


The idea behind a possible speed-up of computation is that the leakage of the excited state from $\mc{Q}$ to $\mc{M}$ is faster, when $\mc{Q}$ interacts with an excited state instead of a ground state. In other words, if we would allow for the excited state to be exchanged between different qubits of $\mc{M}$, then when $\mc{Q}$ starts interacting with the next qubit in $\mc{M}$, the $(k+1)$st qubit in $\mc{M}$ has already a head start.

Motivated by this observation and for the sake of simplicity we thus \emph{design} the interaction between the $N$ qubits in $\mc{M}$ by pairwise SWAP operations,
\begin{equation}
\label{eq:U_t_ancilla}
H^\mc{M}_{k,k+1}=i \Omega\left(\ket{0}\bra{1}_{k}\otimes\ket{1}\bra{0}_{k+1}-\ket{1}\bra{0}_{k}\otimes\ket{0}\bra{1}_{k+1}\right)\,,
\end{equation}
where the strength of interaction is denoted by $\Omega$. Consequently, the total Hamiltonian  of the ``universe'' at time $t$, $k\dt\leq t\leq (k+1)\dt$ consisting of $\mc{Q}$ and $N$ qubits in the quantum memory $\mc{M}$, can be written as
\begin{equation}
H^{\mc{Q}\otimes\mc{M}}=H^\mc{Q}_{k+1}+\sum_{l=1}^{N-1}H^\mc{M}_{l,l+1}\,,
\end{equation}
which is illustrated in Fig.~\ref{p1}. Although still conceptually simple, the Hamiltonian $H^{\mc{Q}\otimes\mc{M}}$ is complicated enough that its dynamics cannot be solved analytically for general $N$ in a closed form. Therefore, we continue with a discussion of numerical findings for $N=3$.
\begin{figure}
\begin{center}
\includegraphics[width=0.48\textwidth]{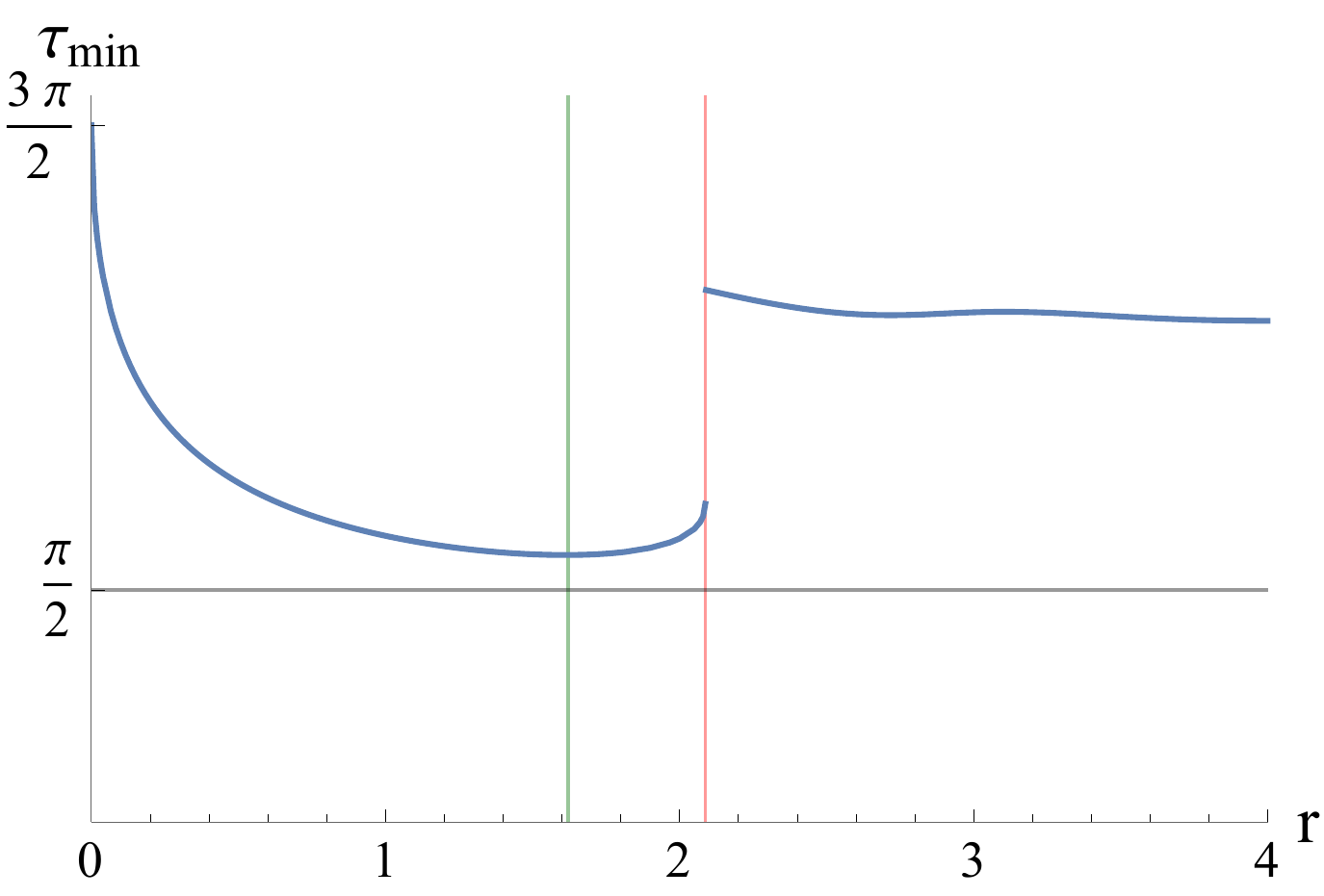}
\caption{Minimal time, $\tau_\mrm{min}$ to complete a qubit flip in $\mc{Q}$ as a function of $r=\Omega/\omega$. The green vertical line marks the optimal ratio $r_{\mathrm{opt}}\approx 1.620000$ that leads to the fastest computation $\tau_{\min}^{(\mathrm{opt})}\approx 1.807022$. The red vertical line denotes a critical point $r_{\mathrm{crit}}\approx 2.087532$, where $\tau_\mrm{min}$ discontinuously increases. Further, for $r\rightarrow \infty$ we see that $\tau_\mrm{min}$ asymptotically approaches $\tau_{\min}^{(r\rightarrow \infty)}\approx 3.332162$. The horizontal line is a reference line that represents the ideal case of a minimal $\tau_\mrm{min}$ for $N=1$.}
\label{fig:tmin_vs_r}
\end{center}
\end{figure}

The first interesting question to ask is whether the minimal time, $\tau_\mrm{min}=3 \Delta t$, necessary to perform a qubit flip, can be minimized as a function of the relative interaction strength $r=\Omega/\omega$. The result of this optimization problem is summarized in Fig.~\ref{fig:tmin_vs_r}.

We observe that $\tau_\mrm{min}$ and thus the optimal lengths of interaction $\Delta t$ strongly depends on $r=\Omega/\omega$. Remarkably, there is a unique minimum, that marks the optimal ratio $r_{\mathrm{opt}}$ at which the qubit flip is performed the fastest. Quite remarkable, the minimal value $\tau_\mrm{min}$ can even become close to the ideal, noninteracting case. In this limit, $\mc{Q}$ interacts with only a single qubit for entire time of interacting, and hence the time for a qubit flip is determined by the ``eigentime", i.e., the time for the unperturbed system to complete one full oscillation in Hilbert space. Thus, for $r_{\mathrm{opt}}$ the quantum Zeno effect is effectively prevented, even though $\mc{Q}$ is frequently interrogated by $\mc{M}$. A second observation is that at a critical value $r_{\mathrm{crit}}$ the minimal time $\tau_\mrm{min}$ jumps to a value which is only slightly better than the worst value given by the case when no interaction between the qubits in $\mc{M}$ is present. In this limit, the rate of interaction between the $N$ qubits in $\mc{M}$ is faster, than the exchange of information of the memory with $\mc{Q}$. Thus, $\mc{Q}$ effectively interacts with the whole memory $\mc{M}$ and no longer qubit-by-qubit.
\begin{figure}
\begin{center}
\includegraphics[width=0.48\hsize]{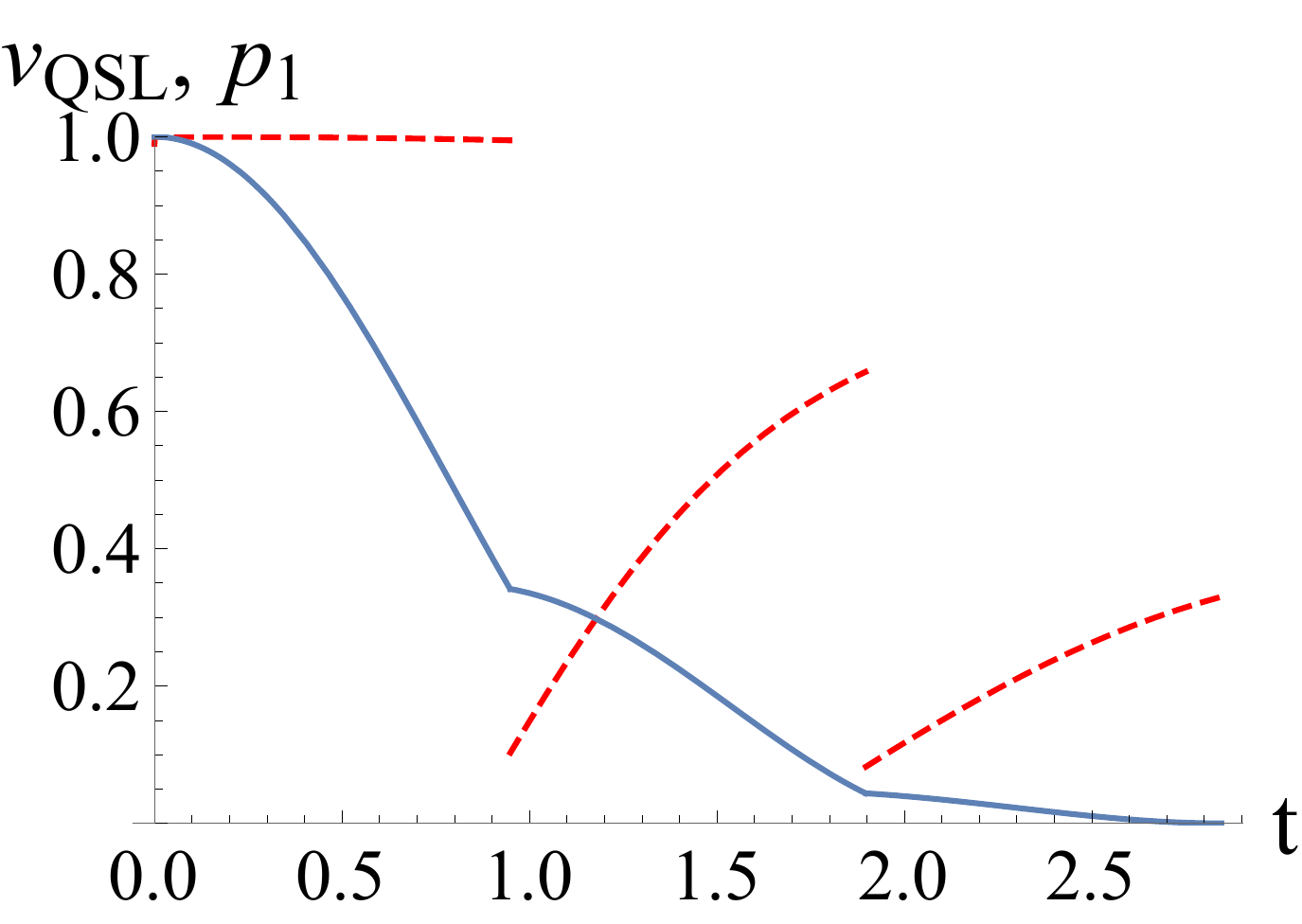}
\includegraphics[width=0.48\hsize]{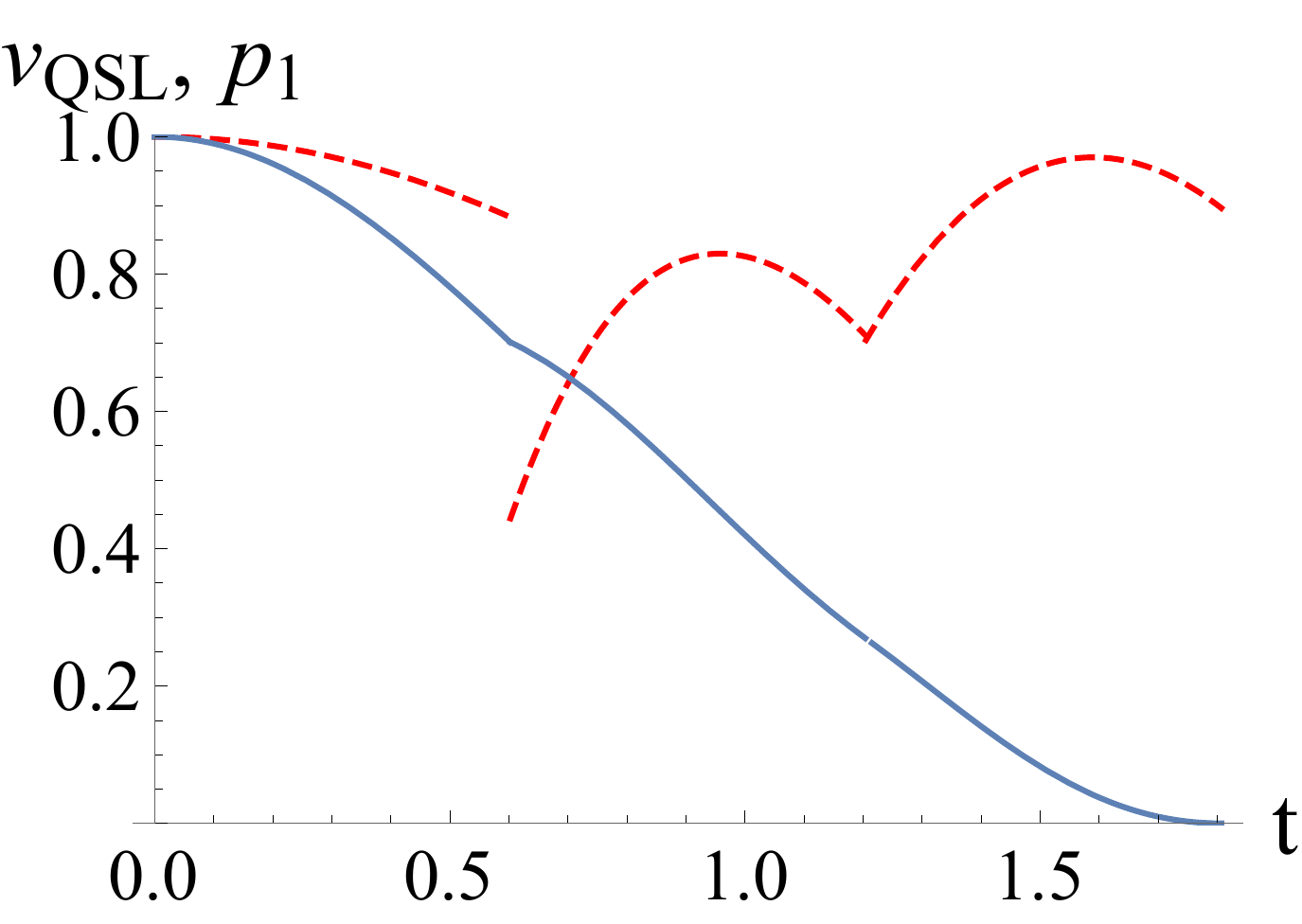}\\
\quad(a)\quad\quad\quad\quad\quad\quad\quad\quad\quad\quad\quad\quad(b)\\
\includegraphics[width=0.48\hsize]{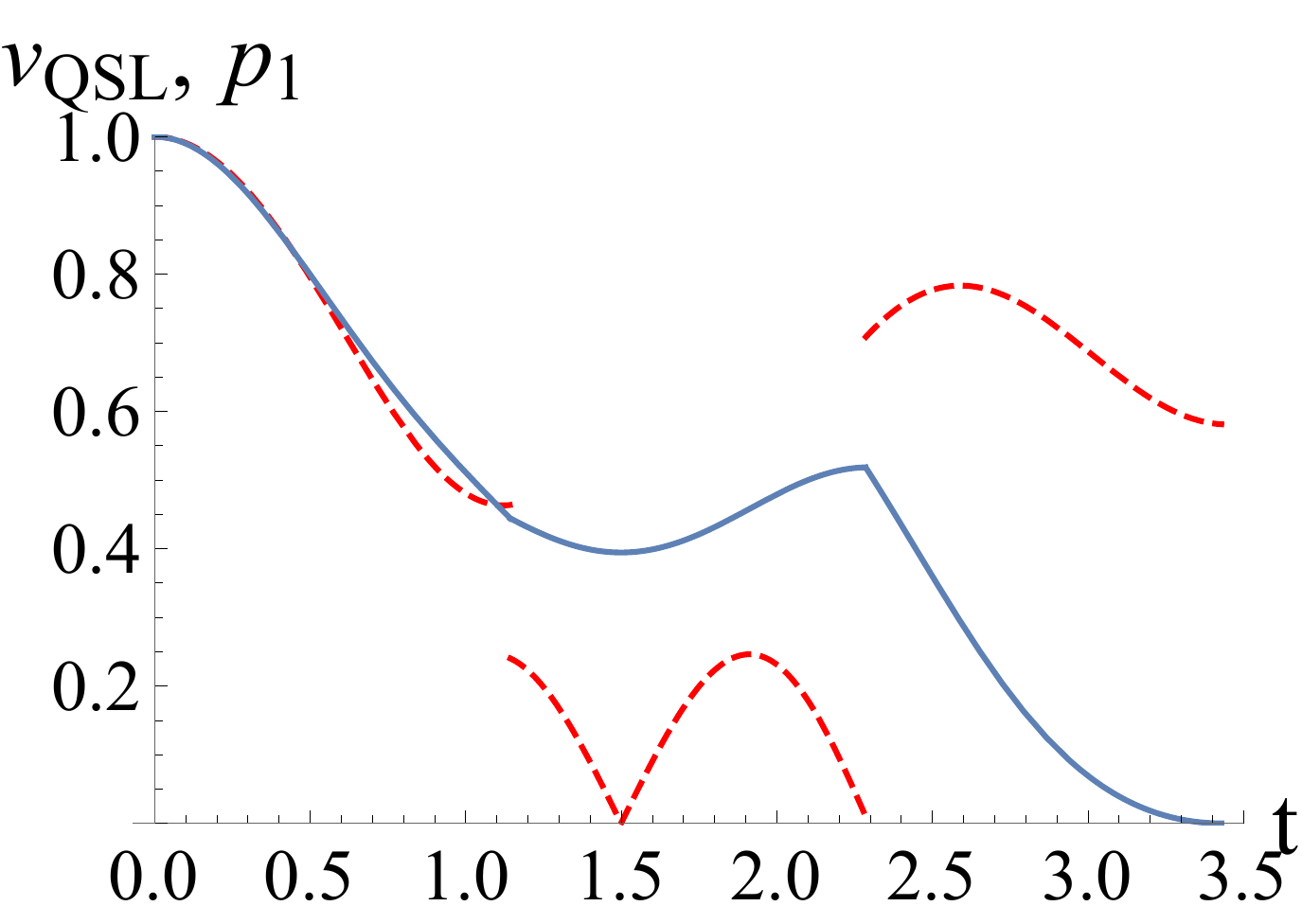}
\includegraphics[width=0.48\hsize]{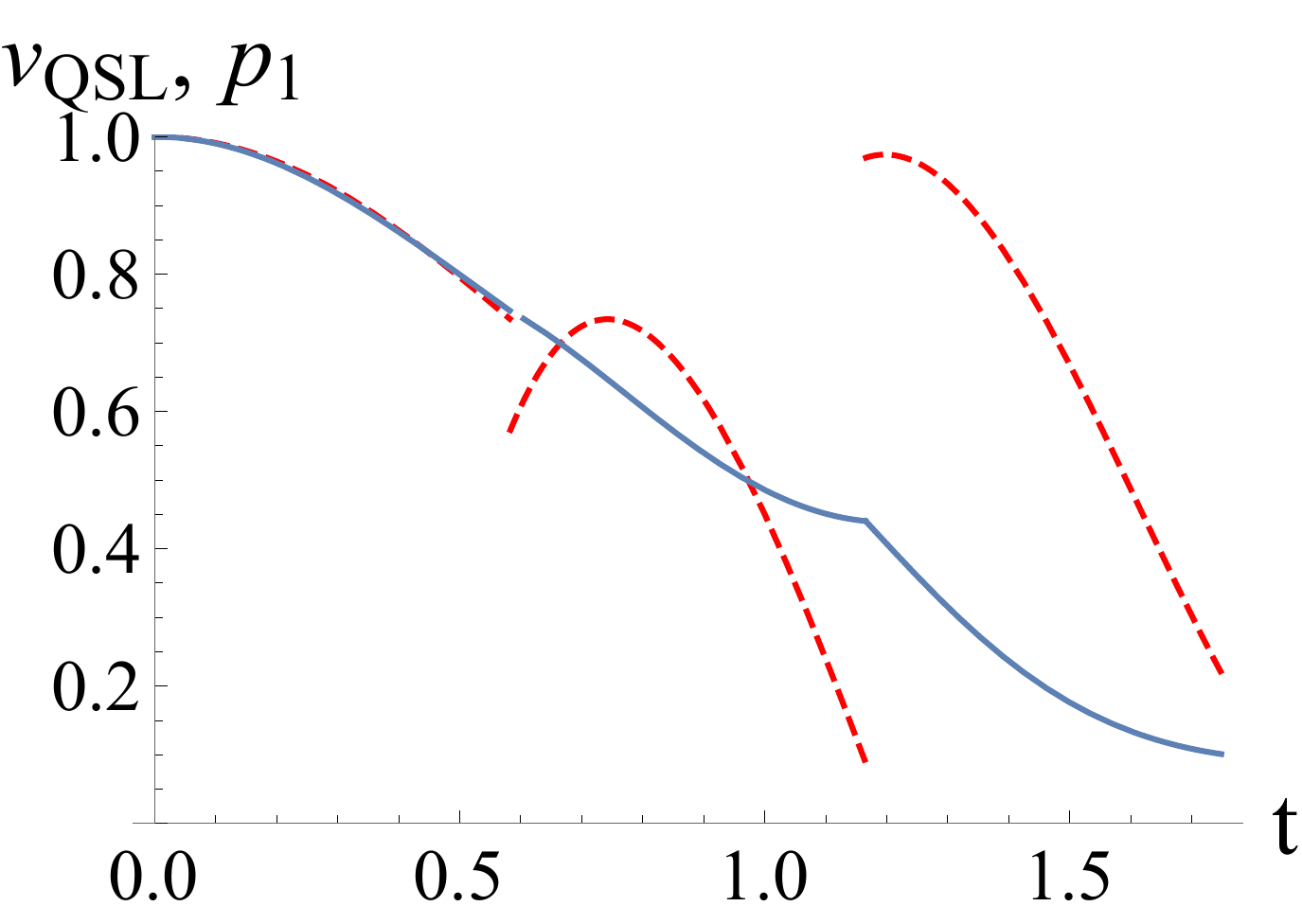}\\
\quad(c)\quad\quad\quad\quad\quad\quad\quad\quad\quad\quad\quad\quad(d)
\caption{Quantum speed limit $v_\mrm{QSL}$ of $\mc{Q}$ \eqref{eq:vqsl} (red, dashed line), and fidelity $p_1(r,t)=\bra{1}\rho_t\ket{1}$ \eqref{eq:reduced_state_iterations} (blue, solid line) for various ratios of interaction strengths $r=\Omega/\omega$ and total times of interaction $\tau$. (a) Small interaction ratio ({$r=0.2$, $\tau=\tau_{\min}$}). (b) Optimal interaction ratio ({$r=r_{\mathrm{opt}}\approx 1.62$, $\tau=\tau_{\min}$}). (c) Super-critical interaction ratio ({$r=2.7>r_{\mathrm{crit}}$, $\tau=\tau_{\min}$}). (d) Super-critical interaction ratio ({$r=2.7$, $\tau$ given by the first local minimum of $p_1(\tau)$}). In (a), (b), and (c), the system qubit is successfully erased at the end of the cycle. However, since the ratio $r$ in (c) is above the critical value, the time it takes to erase is much longer than in $(a)$ or $(b)$. In (d), the fidelity achieves a local minimum at the end of the cycle, but the system qubit fails to be fully erased.}
\label{fig:vsqlVSerasure}
\end{center}
\end{figure}

Finally, we also computed the quantum speed limit $v_\mrm{QSL}(t)$ and the fidelity between instantaneous state $\rho_t$ and target state $\ket{1}$, $p_1(t)=\bra{1}\rho_t\ket{1}$. Results are collected in Fig.~\ref{fig:vsqlVSerasure} for four different values of $r$.  The first graph shows the case of small ratio of interaction $r\ll r_{\mathrm{opt}}$, the second the optimal interaction ratio $r=r_{\mathrm{opt}}$. There, the total time of interaction was chosen as the minimal time of erasure $\tau_{\min}$. Thus, the $\mc{Q}$ is flipped when the interaction stops, $\rho_{\tau_{\min}}=\pro{0}{0}$. The third example shows the case when the ratio is larger than the critical ratio, $r>r_{\mathrm{crit}}$. As a result, $\tau=\tau_{\min}$ is much longer. In the fourth example, we picked the same super-critical ratio, but we chose $\tau$ such that the $p_1(\tau)$ achieves its first local minimum at $\tau$ ($\tau_{\min}$ corresponds to the second local minimum). In this graph, $\mc{Q}$ gets quite close to $\ket{0}$, but fails to be fully erased at the end of the cycle.

\begin{figure}
\begin{center}
\includegraphics[width=.48\textwidth]{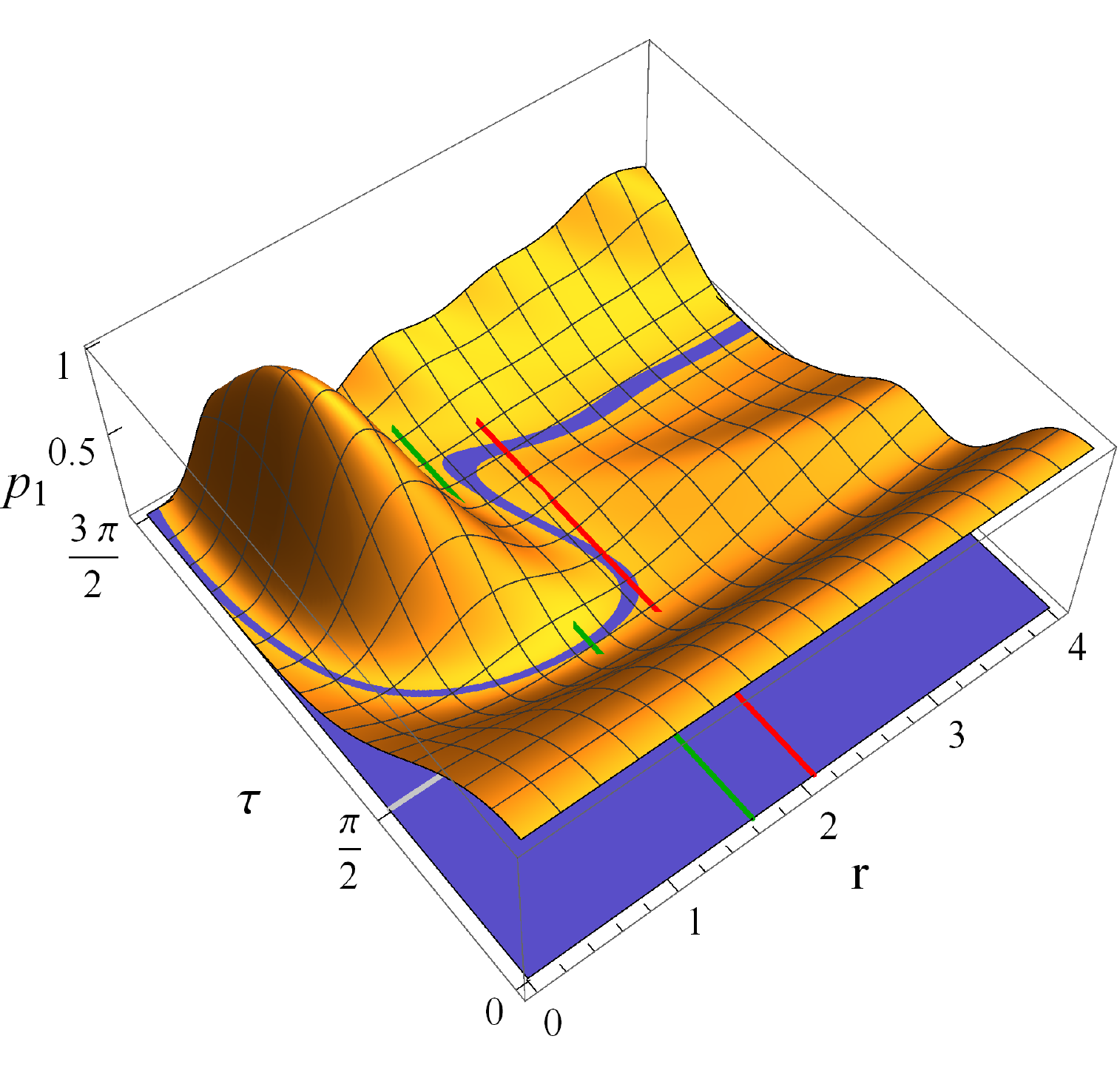}
\caption{Fidelity  $p_1(r,\tau)=\bra{1}\rho_\tau\ket{1}$ as a function of $\tau$ and $r$. The constant function $f=0.001$ (blue) illustrates the minimums of the fidelity. Green and red lines denote the optimal ratio $r_{\mathrm{opt}}$, and the critical ratio $r_{\mathrm{crit}}$ respectively. The horizontal line is a reference line that represents the ideal case of a minimal $\tau_\mrm{min}$ for $N=1$.}
\label{fig:p1}
\end{center}
\end{figure}

To gain further insight into the dynamics of $\mc{Q}$ at $t=\tau$,  we also plot the dependence of the fidelity $p_1(\tau)=\bra{1}\rho_\tau\ket{1}$ as a function of  total time of interaction $\tau$ and ratio of interaction strengths $r$ in Fig.~\ref{fig:p1}. This fairly complicated function shows that only carefully chosen combinations of $\tau$ and $r$ lead to a successful prevention of the quantum Zeno effect, that means here successfully performing a qubit flip.

\paragraph*{Concluding remarks.}

The present analysis revealed that frequent interactions between a computational device $\mc{Q}$ and a quantum hard disk $\mc{M}$ can lead to a failure of computation. In particular, we saw that this ``failure'' is a consequence of the quantum Zeno effect, which is governed by a discontinuity of the quantum speed limit, or more generally by a discontinuity of the quantum Fisher information. Since it was shown that there is a wide class of generic systems for which the Fisher information exhibits discontinuous behavior, our findings are exemplary for what one would expect in more general settings.

In a second part of the analysis we showed that correlations and interaction between the elements of $\mc{M}$ can prevent the failure of computation, but also that there is a trade-off between the interaction strength between the qubits in $\mc{M}$ and the rate with which a computation can be successfully performed. Therefore, we would expect our findings to have profound implication in the design of quantum computers: accessing quantum information frequently by reading into a quantum memory can effectively stall the computation, and lead to errors due to time mismatch. However, carefully designed side interaction could help to sustain the computation.

\acknowledgements{The authors would like to thank the stimulating environment provided by the Telluride Science Research Center, where this project was conceived. D\v{S} acknowledges support from the Foundational
Questions Institute (FQXi.org). SD acknowledges support from the U.S. National Science Foundation under Grant No. CHE-1648973.}

\bibliography{speedbib}

\begin{thebibliography}{37}%
\makeatletter
\providecommand \@ifxundefined [1]{%
 \@ifx{#1\undefined}
}%
\providecommand \@ifnum [1]{%
 \ifnum #1\expandafter \@firstoftwo
 \else \expandafter \@secondoftwo
 \fi
}%
\providecommand \@ifx [1]{%
 \ifx #1\expandafter \@firstoftwo
 \else \expandafter \@secondoftwo
 \fi
}%
\providecommand \natexlab [1]{#1}%
\providecommand \enquote  [1]{``#1''}%
\providecommand \bibnamefont  [1]{#1}%
\providecommand \bibfnamefont [1]{#1}%
\providecommand \citenamefont [1]{#1}%
\providecommand \href@noop [0]{\@secondoftwo}%
\providecommand \href [0]{\begingroup \@sanitize@url \@href}%
\providecommand \@href[1]{\@@startlink{#1}\@@href}%
\providecommand \@@href[1]{\endgroup#1\@@endlink}%
\providecommand \@sanitize@url [0]{\catcode `\\12\catcode `\$12\catcode
  `\&12\catcode `\#12\catcode `\^12\catcode `\_12\catcode `\%12\relax}%
\providecommand \@@startlink[1]{}%
\providecommand \@@endlink[0]{}%
\providecommand \url  [0]{\begingroup\@sanitize@url \@url }%
\providecommand \@url [1]{\endgroup\@href {#1}{\urlprefix }}%
\providecommand \urlprefix  [0]{URL }%
\providecommand \Eprint [0]{\href }%
\providecommand \doibase [0]{http://dx.doi.org/}%
\providecommand \selectlanguage [0]{\@gobble}%
\providecommand \bibinfo  [0]{\@secondoftwo}%
\providecommand \bibfield  [0]{\@secondoftwo}%
\providecommand \translation [1]{[#1]}%
\providecommand \BibitemOpen [0]{}%
\providecommand \bibitemStop [0]{}%
\providecommand \bibitemNoStop [0]{.\EOS\space}%
\providecommand \EOS [0]{\spacefactor3000\relax}%
\providecommand \BibitemShut  [1]{\csname bibitem#1\endcsname}%
\let\auto@bib@innerbib\@empty
\bibitem [{\citenamefont {Sanders}(2017)}]{Sanders2017}%
  \BibitemOpen
  \bibfield  {author} {\bibinfo {author} {\bibfnamefont {B.~C.}\ \bibnamefont
  {Sanders}},\ }\href {\doibase 10.1088/978-0-7503-1536-4} {\emph {\bibinfo
  {title} {How to Build a Quantum Computer}}},\ 2399-2891\ (\bibinfo
  {publisher} {IOP Publishing},\ \bibinfo {year} {2017})\BibitemShut {NoStop}%
\bibitem [{\citenamefont {{Preskill}}()}]{Preskill2012}%
  \BibitemOpen
  \bibfield  {author} {\bibinfo {author} {\bibfnamefont {J.}~\bibnamefont
  {{Preskill}}},\ }\bibfield  {title} {\enquote {\bibinfo {title} {{Quantum
  computing and the entanglement frontier}},}\ }\href
  {https://arxiv.org/abs/1203.5813} {\bibinfo  {journal} {arXiv:1203.5813}\
  }\BibitemShut {NoStop}%
\bibitem [{\citenamefont {Papageorgiou}\ and\ \citenamefont
  {Traub}(2013)}]{Papageorgiou2013}%
  \BibitemOpen
\bibfield  {journal} {  }\bibfield  {author} {\bibinfo {author} {\bibfnamefont
  {A.}~\bibnamefont {Papageorgiou}}\ and\ \bibinfo {author} {\bibfnamefont
  {J.~F.}\ \bibnamefont {Traub}},\ }\bibfield  {title} {\enquote {\bibinfo
  {title} {Measures of quantum computing speedup},}\ }\href {\doibase
  10.1103/PhysRevA.88.022316} {\bibfield  {journal} {\bibinfo  {journal} {Phys.
  Rev. A}\ }\textbf {\bibinfo {volume} {88}},\ \bibinfo {pages} {022316}
  (\bibinfo {year} {2013})}\BibitemShut {NoStop}%
\bibitem [{\citenamefont {Heisenberg}(1927)}]{Heisenberg1927}%
  \BibitemOpen
  \bibfield  {author} {\bibinfo {author} {\bibfnamefont {W.}~\bibnamefont
  {Heisenberg}},\ }\bibfield  {title} {\enquote {\bibinfo {title} {{\"U}ber den
  anschaulichen {I}nhalt der quantentheoretischen {K}inematik und
  {M}echanik},}\ }\href {\doibase 10.1007/BF01397280} {\bibfield  {journal}
  {\bibinfo  {journal} {Z. Phys.}\ }\textbf {\bibinfo {volume} {43}},\ \bibinfo
  {pages} {172} (\bibinfo {year} {1927})}\BibitemShut {NoStop}%
\bibitem [{\citenamefont {Mandelstam}\ and\ \citenamefont
  {Tamm}(1945)}]{Mandelstam1945}%
  \BibitemOpen
  \bibfield  {author} {\bibinfo {author} {\bibfnamefont {L.}~\bibnamefont
  {Mandelstam}}\ and\ \bibinfo {author} {\bibfnamefont {I.}~\bibnamefont
  {Tamm}},\ }\bibfield  {title} {\enquote {\bibinfo {title} {The uncertainty
  relation between energy and time in nonrelativistic quantum mechanics},}\
  }\href {https://link.springer.com/chapter/10.1007/978-3-642-74626-0_8}
  {\bibfield  {journal} {\bibinfo  {journal} {J. Phys.}\ }\textbf {\bibinfo
  {volume} {9}},\ \bibinfo {pages} {249} (\bibinfo {year} {1945})}\BibitemShut
  {NoStop}%
\bibitem [{\citenamefont {Margolus}\ and\ \citenamefont
  {Levitin}(1998)}]{Margolus1998}%
  \BibitemOpen
  \bibfield  {author} {\bibinfo {author} {\bibfnamefont {N.}~\bibnamefont
  {Margolus}}\ and\ \bibinfo {author} {\bibfnamefont {L.~B.}\ \bibnamefont
  {Levitin}},\ }\bibfield  {title} {\enquote {\bibinfo {title} {The maximum
  speed of dynamical evolution},}\ }\href {\doibase
  10.1016/S0167-2789(98)00054-2} {\bibfield  {journal} {\bibinfo  {journal}
  {Physica D}\ }\textbf {\bibinfo {volume} {120}},\ \bibinfo {pages} {188}
  (\bibinfo {year} {1998})}\BibitemShut {NoStop}%
\bibitem [{\citenamefont {Deffner}\ and\ \citenamefont
  {Campbell}(2017)}]{deffner2017quantum}%
  \BibitemOpen
  \bibfield  {author} {\bibinfo {author} {\bibfnamefont {S.}~\bibnamefont
  {Deffner}}\ and\ \bibinfo {author} {\bibfnamefont {S.}~\bibnamefont
  {Campbell}},\ }\bibfield  {title} {\enquote {\bibinfo {title} {Quantum speed
  limits: from {H}eisenberg's uncertainty principle to optimal quantum
  control},}\ }\href
  {http://iopscience.iop.org/article/10.1088/1751-8121/aa86c6/meta} {\bibfield
  {journal} {\bibinfo  {journal} {J. Phys. A: Math. Theor.}\ }\textbf {\bibinfo
  {volume} {50}},\ \bibinfo {pages} {453001} (\bibinfo {year}
  {2017})}\BibitemShut {NoStop}%
\bibitem [{\citenamefont {Feynman}(1982)}]{Feynman1982}%
  \BibitemOpen
  \bibfield  {author} {\bibinfo {author} {\bibfnamefont {R.~P.}\ \bibnamefont
  {Feynman}},\ }\bibfield  {title} {\enquote {\bibinfo {title} {Simulating
  physics with computers},}\ }\href {\doibase 10.1007/BF02650179} {\bibfield
  {journal} {\bibinfo  {journal} {Int. J. Theo. Phys.}\ }\textbf {\bibinfo
  {volume} {21}},\ \bibinfo {pages} {467} (\bibinfo {year} {1982})}\BibitemShut
  {NoStop}%
\bibitem [{\citenamefont {Bremermann}(1967)}]{Bremermann1967}%
  \BibitemOpen
  \bibfield  {author} {\bibinfo {author} {\bibfnamefont {H.~J.}\ \bibnamefont
  {Bremermann}},\ }\bibfield  {title} {\enquote {\bibinfo {title} {Quantum
  noise and information},}\ }in\ \href
  {http://projecteuclid.org/euclid.bsmsp/1200513783} {\emph {\bibinfo
  {booktitle} {Proceedings of the Fifth Berkeley Symposium on Mathematical
  Statistics and Probability, Volume 4: Biology and Problems of Health}}}\
  (\bibinfo  {publisher} {University of California Press},\ \bibinfo {address}
  {Berkeley, Calif.},\ \bibinfo {year} {1967})\ pp.\ \bibinfo {pages}
  {15--20}\BibitemShut {NoStop}%
\bibitem [{\citenamefont {Bekenstein}(1974)}]{Bekenstein1974}%
  \BibitemOpen
  \bibfield  {author} {\bibinfo {author} {\bibfnamefont {J.~D.}\ \bibnamefont
  {Bekenstein}},\ }\bibfield  {title} {\enquote {\bibinfo {title} {Generalized
  second law of thermodynamics in black-hole physics},}\ }\href {\doibase
  10.1103/PhysRevD.9.3292} {\bibfield  {journal} {\bibinfo  {journal} {Phys.
  Rev. D}\ }\textbf {\bibinfo {volume} {9}},\ \bibinfo {pages} {3292} (\bibinfo
  {year} {1974})}\BibitemShut {NoStop}%
\bibitem [{\citenamefont {Bekenstein}\ and\ \citenamefont
  {Schiffer}(1990)}]{Bekenstein1990}%
  \BibitemOpen
  \bibfield  {author} {\bibinfo {author} {\bibfnamefont {J.~D.}\ \bibnamefont
  {Bekenstein}}\ and\ \bibinfo {author} {\bibfnamefont {M.}~\bibnamefont
  {Schiffer}},\ }\bibfield  {title} {\enquote {\bibinfo {title} {Quantum
  limitations on the storage and transmission of information},}\ }\href
  {https://www.worldscientific.com/doi/abs/10.1142/S0129183190000207}
  {\bibfield  {journal} {\bibinfo  {journal} {Int. J. Mod. Phys. C}\ }\textbf
  {\bibinfo {volume} {1}},\ \bibinfo {pages} {355} (\bibinfo {year}
  {1990})}\BibitemShut {NoStop}%
\bibitem [{\citenamefont {Deffner}\ and\ \citenamefont
  {Lutz}(2013)}]{Deffner2013PRL}%
  \BibitemOpen
  \bibfield  {author} {\bibinfo {author} {\bibfnamefont {S.}~\bibnamefont
  {Deffner}}\ and\ \bibinfo {author} {\bibfnamefont {E.}~\bibnamefont {Lutz}},\
  }\bibfield  {title} {\enquote {\bibinfo {title} {Quantum speed limit for
  non-{Markovian} dynamics},}\ }\href {\doibase 10.1103/PhysRevLett.111.010402}
  {\bibfield  {journal} {\bibinfo  {journal} {Phys. Rev. Lett.}\ }\textbf
  {\bibinfo {volume} {111}},\ \bibinfo {pages} {010402} (\bibinfo {year}
  {2013})}\BibitemShut {NoStop}%
\bibitem [{\citenamefont {Cimmarusti}\ \emph {et~al.}(2015)\citenamefont
  {Cimmarusti}, \citenamefont {Yan}, \citenamefont {Patterson}, \citenamefont
  {Corcos}, \citenamefont {Orozco},\ and\ \citenamefont
  {Deffner}}]{Cimmarusti2015}%
  \BibitemOpen
  \bibfield  {author} {\bibinfo {author} {\bibfnamefont {A.~D.}\ \bibnamefont
  {Cimmarusti}}, \bibinfo {author} {\bibfnamefont {Z.}~\bibnamefont {Yan}},
  \bibinfo {author} {\bibfnamefont {B.~D.}\ \bibnamefont {Patterson}}, \bibinfo
  {author} {\bibfnamefont {L.~P.}\ \bibnamefont {Corcos}}, \bibinfo {author}
  {\bibfnamefont {L.~A.}\ \bibnamefont {Orozco}}, \ and\ \bibinfo {author}
  {\bibfnamefont {S.}~\bibnamefont {Deffner}},\ }\bibfield  {title} {\enquote
  {\bibinfo {title} {Environment-assisted speed-up of the field evolution in
  cavity quantum electrodynamics},}\ }\href {\doibase
  10.1103/PhysRevLett.114.233602} {\bibfield  {journal} {\bibinfo  {journal}
  {Phys. Rev. Lett.}\ }\textbf {\bibinfo {volume} {114}},\ \bibinfo {pages}
  {233602} (\bibinfo {year} {2015})}\BibitemShut {NoStop}%
\bibitem [{\citenamefont {Deffner}\ and\ \citenamefont
  {Jarzynski}(2013)}]{Deffner2013PRX}%
  \BibitemOpen
  \bibfield  {author} {\bibinfo {author} {\bibfnamefont {S.}~\bibnamefont
  {Deffner}}\ and\ \bibinfo {author} {\bibfnamefont {C.}~\bibnamefont
  {Jarzynski}},\ }\bibfield  {title} {\enquote {\bibinfo {title} {Information
  processing and the second law of thermodynamics: An inclusive, {H}amiltonian
  approach},}\ }\href {\doibase 10.1103/PhysRevX.3.041003} {\bibfield
  {journal} {\bibinfo  {journal} {Phys. Rev. X}\ }\textbf {\bibinfo {volume}
  {3}},\ \bibinfo {pages} {041003} (\bibinfo {year} {2013})}\BibitemShut
  {NoStop}%
\bibitem [{\citenamefont {Deffner}(2013)}]{Deffner2013}%
  \BibitemOpen
  \bibfield  {author} {\bibinfo {author} {\bibfnamefont {S.}~\bibnamefont
  {Deffner}},\ }\bibfield  {title} {\enquote {\bibinfo {title}
  {Information-driven current in a quantum {Maxwell} demon},}\ }\href {\doibase
  10.1103/PhysRevE.88.062128} {\bibfield  {journal} {\bibinfo  {journal} {Phys.
  Rev. E}\ }\textbf {\bibinfo {volume} {88}},\ \bibinfo {pages} {062128}
  (\bibinfo {year} {2013})}\BibitemShut {NoStop}%
\bibitem [{\citenamefont {Taddei}\ \emph {et~al.}(2013)\citenamefont {Taddei},
  \citenamefont {Escher}, \citenamefont {Davidovich},\ and\ \citenamefont
  {de~Matos~Filho}}]{Taddei2013}%
  \BibitemOpen
  \bibfield  {author} {\bibinfo {author} {\bibfnamefont {M.~M.}\ \bibnamefont
  {Taddei}}, \bibinfo {author} {\bibfnamefont {B.~M.}\ \bibnamefont {Escher}},
  \bibinfo {author} {\bibfnamefont {L.}~\bibnamefont {Davidovich}}, \ and\
  \bibinfo {author} {\bibfnamefont {R.~L.}\ \bibnamefont {de~Matos~Filho}},\
  }\bibfield  {title} {\enquote {\bibinfo {title} {Quantum speed limit for
  physical processes},}\ }\href {\doibase 10.1103/PhysRevLett.110.050402}
  {\bibfield  {journal} {\bibinfo  {journal} {Phys. Rev. Lett.}\ }\textbf
  {\bibinfo {volume} {110}},\ \bibinfo {pages} {050402} (\bibinfo {year}
  {2013})}\BibitemShut {NoStop}%
\bibitem [{\citenamefont {Braunstein}\ and\ \citenamefont
  {Caves}(1994)}]{BraunsteinCaves1994a}%
  \BibitemOpen
  \bibfield  {author} {\bibinfo {author} {\bibfnamefont {S.~L.}\ \bibnamefont
  {Braunstein}}\ and\ \bibinfo {author} {\bibfnamefont {C.~M.}\ \bibnamefont
  {Caves}},\ }\bibfield  {title} {\enquote {\bibinfo {title} {Statistical
  distance and the geometry of quantum states},}\ }\href {\doibase
  10.1103/PhysRevLett.72.3439} {\bibfield  {journal} {\bibinfo  {journal}
  {Phys. Rev. Lett.}\ }\textbf {\bibinfo {volume} {72}},\ \bibinfo {pages}
  {3439} (\bibinfo {year} {1994})}\BibitemShut {NoStop}%
\bibitem [{\citenamefont {Paris}(2009)}]{Paris2009a}%
  \BibitemOpen
  \bibfield  {author} {\bibinfo {author} {\bibfnamefont {M.~G.~A.}\
  \bibnamefont {Paris}},\ }\bibfield  {title} {\enquote {\bibinfo {title}
  {Quantum estimation for quantum technology},}\ }\href {\doibase
  10.1142/S0219749909004839} {\bibfield  {journal} {\bibinfo  {journal} {Int.
  J. Quantum Inf.}\ }\textbf {\bibinfo {volume} {07}},\ \bibinfo {pages} {125}
  (\bibinfo {year} {2009})}\BibitemShut {NoStop}%
\bibitem [{\citenamefont {Szczykulska}\ \emph {et~al.}(2016)\citenamefont
  {Szczykulska}, \citenamefont {Baumgratz},\ and\ \citenamefont
  {Datta}}]{szczykulska2016multi}%
  \BibitemOpen
  \bibfield  {author} {\bibinfo {author} {\bibfnamefont {M.}~\bibnamefont
  {Szczykulska}}, \bibinfo {author} {\bibfnamefont {T.}~\bibnamefont
  {Baumgratz}}, \ and\ \bibinfo {author} {\bibfnamefont {A.}~\bibnamefont
  {Datta}},\ }\bibfield  {title} {\enquote {\bibinfo {title} {Multi-parameter
  quantum metrology},}\ }\href
  {http://dx.doi.org/10.1080/23746149.2016.1230476} {\bibfield  {journal}
  {\bibinfo  {journal} {Adv. Phys. X}\ }\textbf {\bibinfo {volume} {1}},\
  \bibinfo {pages} {621} (\bibinfo {year} {2016})}\BibitemShut {NoStop}%
\bibitem [{\citenamefont {Paraoanu}\ and\ \citenamefont
  {Scutaru}(1998)}]{paraoanu1998bures}%
  \BibitemOpen
  \bibfield  {author} {\bibinfo {author} {\bibfnamefont {G.-S.}\ \bibnamefont
  {Paraoanu}}\ and\ \bibinfo {author} {\bibfnamefont {H.}~\bibnamefont
  {Scutaru}},\ }\bibfield  {title} {\enquote {\bibinfo {title} {Bures distance
  between two displaced thermal states},}\ }\href
  {http://journals.aps.org/pra/abstract/10.1103/PhysRevA.58.869} {\bibfield
  {journal} {\bibinfo  {journal} {Phys. Rev. A}\ }\textbf {\bibinfo {volume}
  {58}},\ \bibinfo {pages} {869} (\bibinfo {year} {1998})}\BibitemShut
  {NoStop}%
\bibitem [{\citenamefont {Zanardi}\ \emph {et~al.}(2007)\citenamefont
  {Zanardi}, \citenamefont {Campos-Venuti},\ and\ \citenamefont
  {Giorda}}]{zanardi2007bures}%
  \BibitemOpen
  \bibfield  {author} {\bibinfo {author} {\bibfnamefont {P.}~\bibnamefont
  {Zanardi}}, \bibinfo {author} {\bibfnamefont {L.}~\bibnamefont
  {Campos-Venuti}}, \ and\ \bibinfo {author} {\bibfnamefont {P.}~\bibnamefont
  {Giorda}},\ }\bibfield  {title} {\enquote {\bibinfo {title} {Bures metric
  over thermal state manifolds and quantum criticality},}\ }\href
  {http://journals.aps.org/pra/abstract/10.1103/PhysRevA.76.062318} {\bibfield
  {journal} {\bibinfo  {journal} {Phys. Rev. A}\ }\textbf {\bibinfo {volume}
  {76}},\ \bibinfo {pages} {062318} (\bibinfo {year} {2007})}\BibitemShut
  {NoStop}%
\bibitem [{\citenamefont {Campos-Venuti}\ and\ \citenamefont
  {Zanardi}(2007)}]{venuti2007quantum}%
  \BibitemOpen
  \bibfield  {author} {\bibinfo {author} {\bibfnamefont {L.}~\bibnamefont
  {Campos-Venuti}}\ and\ \bibinfo {author} {\bibfnamefont {P.}~\bibnamefont
  {Zanardi}},\ }\bibfield  {title} {\enquote {\bibinfo {title} {Quantum
  critical scaling of the geometric tensors},}\ }\href
  {https://journals.aps.org/prl/abstract/10.1103/PhysRevLett.99.095701}
  {\bibfield  {journal} {\bibinfo  {journal} {Phys. Rev. Lett.}\ }\textbf
  {\bibinfo {volume} {99}},\ \bibinfo {pages} {095701} (\bibinfo {year}
  {2007})}\BibitemShut {NoStop}%
\bibitem [{\citenamefont {Gu}(2010)}]{gu2010fidelity}%
  \BibitemOpen
  \bibfield  {author} {\bibinfo {author} {\bibfnamefont {S.-J.}\ \bibnamefont
  {Gu}},\ }\bibfield  {title} {\enquote {\bibinfo {title} {Fidelity approach to
  quantum phase transitions},}\ }\href
  {http://www.worldscientific.com/doi/abs/10.1142/S0217979210056335} {\bibfield
   {journal} {\bibinfo  {journal} {Int. J. Mod. Phys. B}\ }\textbf {\bibinfo
  {volume} {24}},\ \bibinfo {pages} {4371} (\bibinfo {year}
  {2010})}\BibitemShut {NoStop}%
\bibitem [{\citenamefont {Banchi}\ \emph {et~al.}(2014)\citenamefont {Banchi},
  \citenamefont {Giorda},\ and\ \citenamefont {Zanardi}}]{banchi2014quantum}%
  \BibitemOpen
  \bibfield  {author} {\bibinfo {author} {\bibfnamefont {L.}~\bibnamefont
  {Banchi}}, \bibinfo {author} {\bibfnamefont {P.}~\bibnamefont {Giorda}}, \
  and\ \bibinfo {author} {\bibfnamefont {P.}~\bibnamefont {Zanardi}},\
  }\bibfield  {title} {\enquote {\bibinfo {title} {Quantum information-geometry
  of dissipative quantum phase transitions},}\ }\href
  {http://journals.aps.org/pre/abstract/10.1103/PhysRevE.89.022102} {\bibfield
  {journal} {\bibinfo  {journal} {Phys. Rev. E}\ }\textbf {\bibinfo {volume}
  {89}},\ \bibinfo {pages} {022102} (\bibinfo {year} {2014})}\BibitemShut
  {NoStop}%
\bibitem [{\citenamefont {Wu}\ and\ \citenamefont
  {Xu}(2016)}]{wu2016geometric}%
  \BibitemOpen
  \bibfield  {author} {\bibinfo {author} {\bibfnamefont {W.}~\bibnamefont
  {Wu}}\ and\ \bibinfo {author} {\bibfnamefont {J.-B.}\ \bibnamefont {Xu}},\
  }\bibfield  {title} {\enquote {\bibinfo {title} {Geometric phase, quantum
  {Fisher} information, geometric quantum correlation and quantum phase
  transition in the cavity-{Bose}-{Einstein}-condensate system},}\ }\href
  {http://link.springer.com/article/10.1007/s11128-015-1186-7} {\bibfield
  {journal} {\bibinfo  {journal} {QIP}\ }\textbf {\bibinfo {volume} {15}},\
  \bibinfo {pages} {3695} (\bibinfo {year} {2016})}\BibitemShut {NoStop}%
\bibitem [{\citenamefont {Marzolino}\ and\ \citenamefont
  {Prosen}(2017)}]{marzolino2017fisher}%
  \BibitemOpen
  \bibfield  {author} {\bibinfo {author} {\bibfnamefont {U.}~\bibnamefont
  {Marzolino}}\ and\ \bibinfo {author} {\bibfnamefont {T.}~\bibnamefont
  {Prosen}},\ }\bibfield  {title} {\enquote {\bibinfo {title} {Fisher
  information approach to nonequilibrium phase transitions in a quantum {XXZ}
  spin chain with boundary noise},}\ }\href
  {https://journals.aps.org/prb/abstract/10.1103/PhysRevB.96.104402} {\bibfield
   {journal} {\bibinfo  {journal} {Phys. Rev. B}\ }\textbf {\bibinfo {volume}
  {96}},\ \bibinfo {pages} {104402} (\bibinfo {year} {2017})}\BibitemShut
  {NoStop}%
\bibitem [{\citenamefont {Hauke}\ \emph {et~al.}(2016)\citenamefont {Hauke},
  \citenamefont {Heyl}, \citenamefont {Tagliacozzo},\ and\ \citenamefont
  {Zoller}}]{hauke2016measuring}%
  \BibitemOpen
  \bibfield  {author} {\bibinfo {author} {\bibfnamefont {P.}~\bibnamefont
  {Hauke}}, \bibinfo {author} {\bibfnamefont {M.}~\bibnamefont {Heyl}},
  \bibinfo {author} {\bibfnamefont {L.}~\bibnamefont {Tagliacozzo}}, \ and\
  \bibinfo {author} {\bibfnamefont {P.}~\bibnamefont {Zoller}},\ }\bibfield
  {title} {\enquote {\bibinfo {title} {Measuring multipartite entanglement
  through dynamic susceptibilities},}\ }\href
  {https://www.nature.com/articles/nphys3700} {\bibfield  {journal} {\bibinfo
  {journal} {Nat. Phys.}\ }\textbf {\bibinfo {volume} {12}},\ \bibinfo {pages}
  {778} (\bibinfo {year} {2016})}\BibitemShut {NoStop}%
\bibitem [{\citenamefont {Girolami}()}]{girolami2017information}%
  \BibitemOpen
  \bibfield  {author} {\bibinfo {author} {\bibfnamefont {D.}~\bibnamefont
  {Girolami}},\ }\bibfield  {title} {\enquote {\bibinfo {title} {Information
  geometry of quantum resources},}\ }\href {https://arxiv.org/abs/1709.05531}
  {\bibinfo  {journal} {arXiv:1709.05531}\ }\BibitemShut {NoStop}%
\bibitem [{\citenamefont {Liu}\ \emph {et~al.}(2017)\citenamefont {Liu},
  \citenamefont {Wang}, \citenamefont {Sun},\ and\ \citenamefont
  {Ye}}]{liu2017quantum}%
  \BibitemOpen
\bibfield  {journal} {  }\bibfield  {author} {\bibinfo {author} {\bibfnamefont
  {C.-C.}\ \bibnamefont {Liu}}, \bibinfo {author} {\bibfnamefont
  {D.}~\bibnamefont {Wang}}, \bibinfo {author} {\bibfnamefont {W.-Y.}\
  \bibnamefont {Sun}}, \ and\ \bibinfo {author} {\bibfnamefont
  {L.}~\bibnamefont {Ye}},\ }\bibfield  {title} {\enquote {\bibinfo {title}
  {Quantum {Fisher} information, quantum entanglement and correlation close to
  quantum critical phenomena},}\ }\href
  {https://link.springer.com/article/10.1007/s11128-017-1674-z} {\bibfield
  {journal} {\bibinfo  {journal} {QIP}\ }\textbf {\bibinfo {volume} {16}},\
  \bibinfo {pages} {219} (\bibinfo {year} {2017})}\BibitemShut {NoStop}%
\bibitem [{\citenamefont {Mancino}\ \emph {et~al.}()\citenamefont {Mancino},
  \citenamefont {Cavina}, \citenamefont {De~Pasquale}, \citenamefont
  {Sbroscia}, \citenamefont {Booth}, \citenamefont {Roccia}, \citenamefont
  {Gianani}, \citenamefont {Giovannetti},\ and\ \citenamefont
  {Barbieri}}]{mancino2018geometrical}%
  \BibitemOpen
  \bibfield  {author} {\bibinfo {author} {\bibfnamefont {L.}~\bibnamefont
  {Mancino}}, \bibinfo {author} {\bibfnamefont {V.}~\bibnamefont {Cavina}},
  \bibinfo {author} {\bibfnamefont {A.}~\bibnamefont {De~Pasquale}}, \bibinfo
  {author} {\bibfnamefont {M.}~\bibnamefont {Sbroscia}}, \bibinfo {author}
  {\bibfnamefont {R.~I.}\ \bibnamefont {Booth}}, \bibinfo {author}
  {\bibfnamefont {E.}~\bibnamefont {Roccia}}, \bibinfo {author} {\bibfnamefont
  {I.}~\bibnamefont {Gianani}}, \bibinfo {author} {\bibfnamefont
  {V.}~\bibnamefont {Giovannetti}}, \ and\ \bibinfo {author} {\bibfnamefont
  {M.}~\bibnamefont {Barbieri}},\ }\bibfield  {title} {\enquote {\bibinfo
  {title} {Geometrical bounds on irreversibility in open quantum systems},}\
  }\href {https://arxiv.org/abs/1801.05188} {\bibinfo  {journal}
  {arXiv:1801.05188}\ }\BibitemShut {NoStop}%
\bibitem [{\citenamefont {Campbell}\ \emph {et~al.}(2018)\citenamefont
  {Campbell}, \citenamefont {Genoni},\ and\ \citenamefont
  {Deffner}}]{Campbell2018}%
  \BibitemOpen
\bibfield  {journal} {  }\bibfield  {author} {\bibinfo {author} {\bibfnamefont
  {S.}~\bibnamefont {Campbell}}, \bibinfo {author} {\bibfnamefont {M.~G.}\
  \bibnamefont {Genoni}}, \ and\ \bibinfo {author} {\bibfnamefont
  {S.}~\bibnamefont {Deffner}},\ }\bibfield  {title} {\enquote {\bibinfo
  {title} {Precision thermometry and the quantum speed limit},}\ }\href
  {http://stacks.iop.org/2058-9565/3/i=2/a=025002} {\bibfield  {journal}
  {\bibinfo  {journal} {Quant. Sci. Technol.}\ }\textbf {\bibinfo {volume}
  {3}},\ \bibinfo {pages} {025002} (\bibinfo {year} {2018})}\BibitemShut
  {NoStop}%
\bibitem [{\citenamefont
  {{\v{S}}afr{\'a}nek}(2017)}]{vsafranek2017discontinuities}%
  \BibitemOpen
  \bibfield  {author} {\bibinfo {author} {\bibfnamefont {D.}~\bibnamefont
  {{\v{S}}afr{\'a}nek}},\ }\bibfield  {title} {\enquote {\bibinfo {title}
  {Discontinuities of the quantum {Fisher information and the Bures} metric},}\
  }\href {https://journals.aps.org/pra/abstract/10.1103/PhysRevA.95.052320}
  {\bibfield  {journal} {\bibinfo  {journal} {Phys. Rev. A}\ }\textbf {\bibinfo
  {volume} {95}},\ \bibinfo {pages} {052320} (\bibinfo {year}
  {2017})}\BibitemShut {NoStop}%
\bibitem [{\citenamefont {Misra}\ and\ \citenamefont
  {Sudarshan}(1977)}]{Misra1977}%
  \BibitemOpen
  \bibfield  {author} {\bibinfo {author} {\bibfnamefont {B.}~\bibnamefont
  {Misra}}\ and\ \bibinfo {author} {\bibfnamefont {E.~C.~G.}\ \bibnamefont
  {Sudarshan}},\ }\bibfield  {title} {\enquote {\bibinfo {title} {The {Zeno}'s
  paradox in quantum theory},}\ }\href {\doibase 10.1063/1.523304} {\bibfield
  {journal} {\bibinfo  {journal} {J. Math. Phys.}\ }\textbf {\bibinfo {volume}
  {18}},\ \bibinfo {pages} {756--763} (\bibinfo {year} {1977})}\BibitemShut
  {NoStop}%
\bibitem [{\citenamefont {Schieve}\ \emph {et~al.}(1989)\citenamefont
  {Schieve}, \citenamefont {Horwitz},\ and\ \citenamefont
  {Levitan}}]{schieve1989numerical}%
  \BibitemOpen
  \bibfield  {author} {\bibinfo {author} {\bibfnamefont {W.~C.}\ \bibnamefont
  {Schieve}}, \bibinfo {author} {\bibfnamefont {L.~P.}\ \bibnamefont
  {Horwitz}}, \ and\ \bibinfo {author} {\bibfnamefont {J.}~\bibnamefont
  {Levitan}},\ }\bibfield  {title} {\enquote {\bibinfo {title} {Numerical study
  of zeno and anti-zeno effects in a local potential model},}\ }\href
  {https://www.sciencedirect.com/science/article/pii/0375960189908116}
  {\bibfield  {journal} {\bibinfo  {journal} {Phys. Lett. A}\ }\textbf
  {\bibinfo {volume} {136}},\ \bibinfo {pages} {264} (\bibinfo {year}
  {1989})}\BibitemShut {NoStop}%
\bibitem [{\citenamefont {Kofman}\ and\ \citenamefont
  {Kurizki}(2000)}]{kofman2000acceleration}%
  \BibitemOpen
  \bibfield  {author} {\bibinfo {author} {\bibfnamefont {A.~G.}\ \bibnamefont
  {Kofman}}\ and\ \bibinfo {author} {\bibfnamefont {G.}~\bibnamefont
  {Kurizki}},\ }\bibfield  {title} {\enquote {\bibinfo {title} {Acceleration of
  quantum decay processes by frequent observations},}\ }\href
  {https://www.nature.com/articles/35014537} {\bibfield  {journal} {\bibinfo
  {journal} {Nature}\ }\textbf {\bibinfo {volume} {405}},\ \bibinfo {pages}
  {546} (\bibinfo {year} {2000})}\BibitemShut {NoStop}%
\bibitem [{\citenamefont {Facchi}\ \emph {et~al.}(2001)\citenamefont {Facchi},
  \citenamefont {Nakazato},\ and\ \citenamefont
  {Pascazio}}]{facchi2001quantum}%
  \BibitemOpen
  \bibfield  {author} {\bibinfo {author} {\bibfnamefont {P.}~\bibnamefont
  {Facchi}}, \bibinfo {author} {\bibfnamefont {H.}~\bibnamefont {Nakazato}}, \
  and\ \bibinfo {author} {\bibfnamefont {S.}~\bibnamefont {Pascazio}},\
  }\bibfield  {title} {\enquote {\bibinfo {title} {From the quantum {Zeno} to
  the inverse quantum {Zeno} effect},}\ }\href
  {https://journals.aps.org/prl/abstract/10.1103/PhysRevLett.86.2699}
  {\bibfield  {journal} {\bibinfo  {journal} {Phys. Rev. Lett.}\ }\textbf
  {\bibinfo {volume} {86}},\ \bibinfo {pages} {2699} (\bibinfo {year}
  {2001})}\BibitemShut {NoStop}%
\bibitem [{\citenamefont {Fischer}\ \emph {et~al.}(2001)\citenamefont
  {Fischer}, \citenamefont {Guti{\'e}rrez-Medina},\ and\ \citenamefont
  {Raizen}}]{fischer2001observation}%
  \BibitemOpen
  \bibfield  {author} {\bibinfo {author} {\bibfnamefont {M.~C.}\ \bibnamefont
  {Fischer}}, \bibinfo {author} {\bibfnamefont {B.}~\bibnamefont
  {Guti{\'e}rrez-Medina}}, \ and\ \bibinfo {author} {\bibfnamefont {M.~G.}\
  \bibnamefont {Raizen}},\ }\bibfield  {title} {\enquote {\bibinfo {title}
  {Observation of the quantum {Zeno and anti-Zeno} effects in an unstable
  system},}\ }\href
  {https://journals.aps.org/prl/abstract/10.1103/PhysRevLett.87.040402}
  {\bibfield  {journal} {\bibinfo  {journal} {Phys. Rev. Lett.}\ }\textbf
  {\bibinfo {volume} {87}},\ \bibinfo {pages} {040402} (\bibinfo {year}
  {2001})}\BibitemShut {NoStop}%
\end{thebibliography}%

\end{document}